# Dressed excitonic states and quantum interference in a three-level quantum dot ladder system

B. D. Gerardot[1], D. Brunner[1], P. A. Dalgarno[1], K. Karrai[2], A. Badolato[3], P. M. Petroff[3], and R. J. Warburton[1]

[1]*School of Engineering and Physical Sciences, Heriot-Watt University, Edinburgh EH14 4AS, UK*
[2]*Center for NanoScience and Department für Physik der LMU, Geschwister-Scholl-Platz 1, 80539 Munich, Germany*
[3]*Materials Department, University of California, Santa Barbara, California 93106*
email: b.d.gerardot@hw.ac.uk

We observe dressed states and quantum interference effects in a strongly driven three-level quantum dot ladder system. The effect of a strong coupling field on one dipole transition is measured by a weak probe field on the second dipole transition using differential reflection. When the coupling energy is much larger than both the homogeneous and inhomogeneous linewidths an Autler-Townes splitting is observed. Striking differences are observed when the transitions resonant with the strong and weak fields are swapped, particularly when the coupling energy is nearly equal to the measured linewidth. This result is attributed to quantum interference: a modest destructive or constructive interference is observed depending on the pump / probe geometry. The data demonstrate that coherence of both the bi-exciton and the exciton is maintained in this solid-state system, even under intense illumination, which is crucial for prospects in quantum information processing and non-linear optical devices.

Strong light-matter coupling of a two-level atom produces a coherent evolution of the atomic state populations, referred to as Rabi flopping. This coherence can be extended to a strongly driven three-level atom, where striking phenomena such as Autler-Townes splitting, dark states, and electromagnetic induced transparency (EIT) can be observed [1]. At the heart of dramatic effects such as EIT is quantum interference where coherence of the driving field and the individual atomic states is crucial.

In recent years, several experiments have proven the atom-like properties of self-assembled quantum dots (QDs). Significantly, the coherence of the ground state (|1>) to exciton (|2>) transition has been explored in neutral [2-4] and negatively charged [5] QDs. However, the coherent properties of a driven three-level ladder QD system are also highly relevant [6, 7]. The bi-exciton (|3>) to |2> to |1> cascade in QDs is particularly interesting due to the ability to generate entangled photon pairs [8-10] and construct a two-bit quantum gate [11]. For solid-state media, a significant issue is whether or not dephasing mechanisms are sufficiently suppressed for quantum interference effects to be manifest. In addition to spontaneous emission, coupling of the discrete quantum states to a continuum of states with uncontrolled degrees of freedom can lead to detrimental dephasing. Examples of deleterious coupling mechanisms include tunnelling, phonon interaction via spin-orbit coupling, hyperfine interaction, and many-body interactions under intense driving fields. Here we perform resonant pump and probe spectroscopy on a single QD ladder system. We observe the dressed states of each QD transition and demonstrate that coherence in this solid-state system is maintained under intense driving fields. Furthermore, evidence of modest quantum interference effects is elicited by swapping the pump and probe fields. In fact, the nature of the quantum interference changes from destructive to constructive depending on the pump / probe geometry.

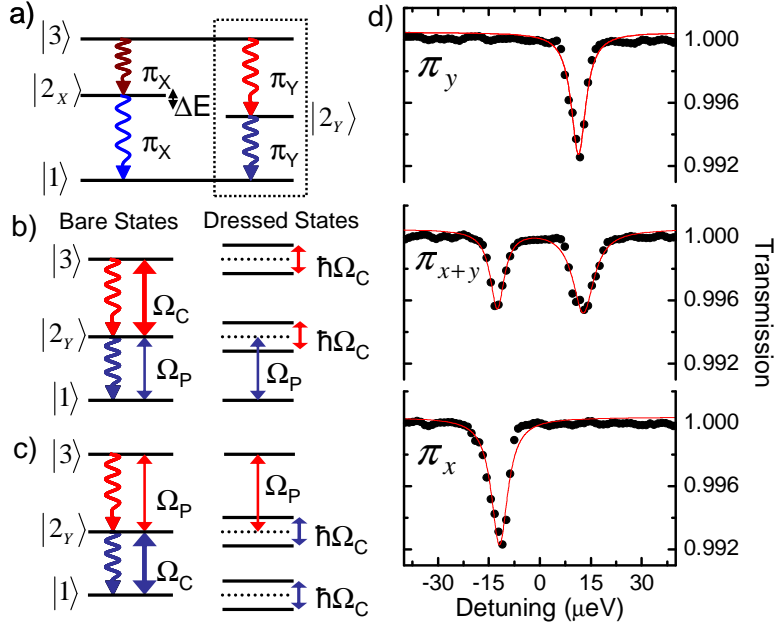

Figure 1: (a) Schematic representation of the QD *s*-shell 4-level system. With pure $\pi_y$ polarization, a three level ladder system is obtained (dashed box). (b) In the first experiment, a strong driving field, $\Omega_C$, is applied to $|2_Y>$ - $|3>$ while a perturbative probe, $\Omega_P$, is scanned over $|1>$ - $|2_Y>$. When $\hbar\Omega_C > \hbar\gamma_{32}$ the dressed state picture is appropriate (right hand side). (c) In the second experiment $\Omega_C$ is applied to $|1>$ - $|2_Y>$ and $\Omega_P$ to $|2_Y>$ - $|3>$. (d) Transmission spectra as $\Omega_P$ is scanned over the $|1>$ - $|2>$ transitions using three different linear polarizations. Here $\hbar\Omega_C = 0$ and the solid lines are Lorentzian fits to the data.

The QD s-shell level schematic is shown in Figure 1a. Due to the electron-hole exchange interaction, the neutral exciton exhibits a fine-structure with two linearly polarized ($\pi_x$ and $\pi_y$) transitions [12], energetically split for the QD studied in this report by 25 μeV (Fig. 1d). Spontaneous emission leads to homogeneous linewidths $\hbar\gamma_{32}$ and $\hbar\gamma_{21}$. In this QD, the bi-exciton is red-shifted by 3.2 meV from the single exciton due to excitonic Coulomb interaction. We obtain a three-level ladder system by choosing to work in the $\pi_Y$ basis (dashed area Fig. 1a). To explore the coherence in the system, we apply a strong coupling field with energy $\hbar\Omega_C$ resonant with either the $|2_Y>$ -$|3>$ or $|1>$ - $|2_Y>$ transition and a weak probe field with energy $\hbar\Omega_P$ resonant with the other transition (Fig.

1b, c). For $\hbar\Omega_C > \hbar\gamma$, a perturbative description of the system using Fermi's golden rule fails and the dressed state picture, which admixes the photon and exciton eigenstates, is appropriate. In the dressed state picture, the bare states are split by $\hbar\Omega_C$ (Fig. 1b and c). As the probe beam is detuned relative to the bare transition two Lorentzian resonances are present: the Autler-Townes doublet [13].

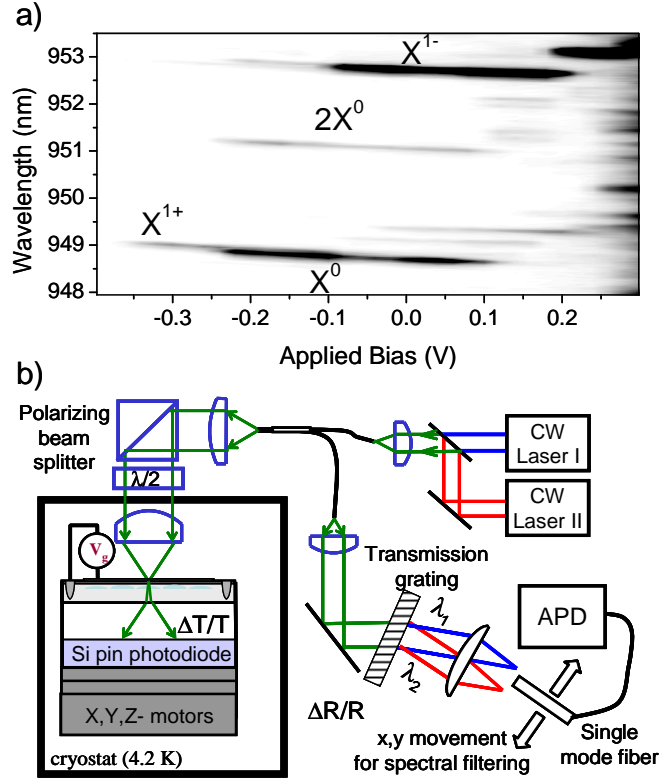

Figure 2: (a) Photoluminescence spectra as a function of applied voltage. The bi-exciton ($2X^0$) is redshifted from the single exciton ($X^0$) by 3.2 meV. For the resonant experiments, the DC-Stark shift is used to detune the QD states relative to the laser energy. The data presented in Figs. 3, 4, and 5 were taken with $V_g \approx -0.15$ V. (b) For the experimental setup, two tunable external cavity diode lasers are coupled into a single mode fiber and focused onto the QD sample after passing through a polarizing beam splitter and half-waveplate. Differential transmission is measured *in situ*. To filter out the strong coupling field, a single mode fiber is spatially positioned to collect only the probe field after the reflection signal passes through a transmission grating. The probe absorption signal is measured with an avalanche photodiode.

Our sample consists of self-assembled InAs / GaAs quantum dots embedded in a charge-tunable heterostructure. We can dictate the charge state of a single QD by the applied bias [14]. The sample used is the same as in ref. 15. Using a confocal microscope, we first characterize a QD using photoluminescence (Fig. 2a) before switching to resonant laser spectroscopy. For this QD, we find identical linear DC Stark shifts as a function of applied bias over the extent of the voltage plateau for both the bi-exciton and exciton states ($1.14 \pm 0.05$ meV/V). We can detect the differential forward scattered signal ($\Delta R/R$) outside of the cryostat [16] or backscattered signal ($\Delta T/T$) *in situ* [17]. The single exciton transition is first characterized in transmission (Fig. 1d). The QD examined here shows linewidths ranging from ~ 1.8 to 4.5 μeV depending on the experimental measurement time. We observe that fast measurement (time constant = 5 ms) yields the smallest linewidths and slow measurement (time constant $\geq 0.2$ s) yields the largest linewidths, consistent with the picture of inhomogeneous broadening due to spectral fluctuations [18 *Supplementary Information*]. Direct lifetime ($\tau$) measurements on many similar QDs yield statistics exhibiting a ratio of $0.65 \pm 0.1$ for $\tau_{32}/\tau_{21}$ and typical values for $\hbar\gamma_{32y}$ and $\hbar\gamma_{21}$ are 0.74 and 1.13 μeV, respectively [19]. In the transmission geometry, both the pump and probe beams strike the detector and the pump laser shot noise overwhelms the probe laser signal. In fact, the noise equivalent power is ~ $10^4$ times worse for a strong driving field compared to a weak field [15]. Therefore, to perform the two-colour pump / probe experiment we measure in reflection and filter out the strong driving field with greater than $10^3$ extinction ratio (Fig. 2b). In this way we can measure the probe signal with high signal:noise. We note that differential transmission measurements yield Lorentzian lineshapes while differential reflection lineshapes have a dispersive component. This is due to an interference effect: the highly coherent laser interacts with a cavity formed between the sample surface and polished fibre tip [see ref. 16 for a study of this interference effect with a shorter cavity length]. This interaction varies as a function of photon energy, hence the lineshapes in Figs. 3 and 5 are slightly asymmetric. We note that the absence of any asymmetry or overshoot in the lineshapes observed in the transmission geometry under strong excitation rules out the presence of a Fano effect in the heterostructure [20]. Hence, dephasing due to coherent coupling with nearby continuum states is sufficiently suppressed in this sample.

Figure 3a shows results for driving the $|2_Y\rangle$ - $|3\rangle$ transition on resonance with $\Omega_C$ and probing the $|1\rangle$ - $|2_Y\rangle$ transition with $\Omega_P$. As $\hbar\Omega_C$ is increased from 0, the single peak splits into two. This splitting is directly proportional to the amplitude of the coupling field (as shown in Fig. 4), consistent with the Autler-Townes splitting. In this experiment, a maximum coupling field power of 100 µW was used to generate a peak to peak energy splitting of 67 µeV. Using the 4-level model described below, we find that the peak to peak splitting is equal to $0.71\hbar\Omega_C$ rather than equal to $\hbar\Omega_C$ for this experiment due to the fact that both $\Omega_C$ and $\Omega_P$ are detuned together using the DC Stark shift, as opposed to the prototypical experiment of detuning only $\Omega_P$. We have therefore achieved $\hbar\Omega_C \cong 100$ µeV, which corresponds to a Rabi flopping period of ~ 6.5 ps. Fig. 3c shows the result of detuning $\Omega_C$ from resonance with $|2_Y\rangle$ - $|3\rangle$ with $\hbar\Omega_C = 24.5$ µeV. An anti-crossing is clearly observed here. Again, the peak to peak splitting is not quite the traditional $(\hbar\delta_C^2 + \hbar\Omega_C^2)^{1/2}$, where $\hbar\delta_C$ is the coupling field detuning energy, due to the fact that both the lasers are detuned simultaneously by the DC Stark shift.

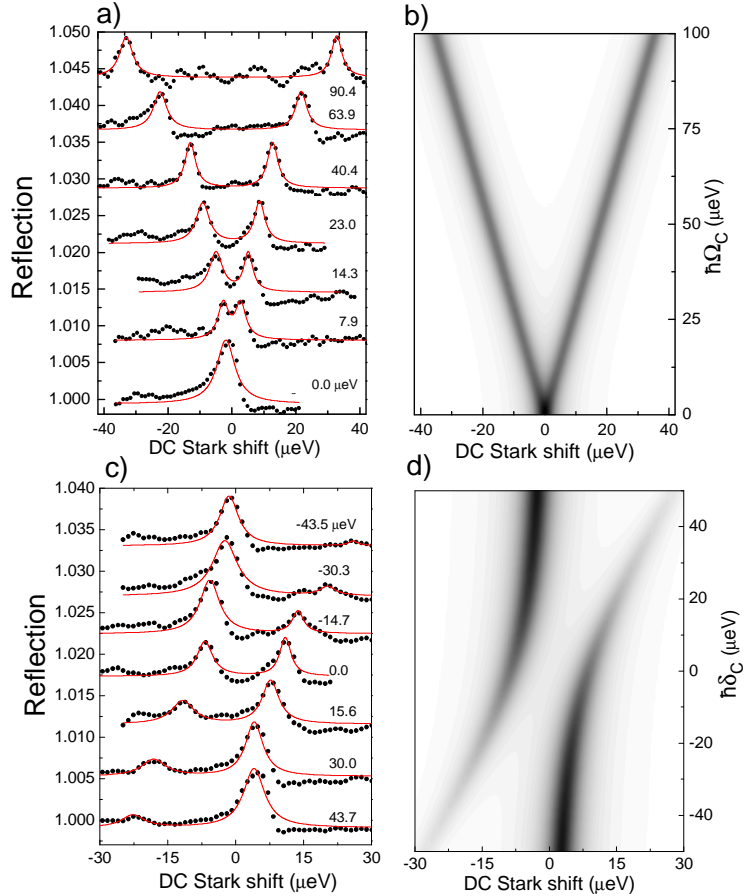

Figure 3: The effect of a coupling field on the probe absorption spectrum. (a) The coupling field is resonant with the $|2_Y\rangle$ - $|3\rangle$ transition for a DC Stark shift of 0 µeV. The peak to peak splitting increases with increasing coupling field amplitude. Each spectrum is offset for clarity. (b) A simulation of the 4 level model using $\gamma_{32y} = 0.74$ µeV, $\gamma_{21} = 1.13$ µeV, $\hbar\Omega_P = 0.4$ µeV, and $\alpha_0 = 0.03$ as a function of $\hbar\Omega_C$. Black (white) colouring corresponds to a signal contrast of 0.007 (0) and the signal is convoluted with a 3 µeV FWHM Lorentzian. (c) The coupling field ($\hbar\Omega_C = 24.5$ µeV) is detuned relative to the $|2_Y\rangle$ - $|3\rangle$ transition. A simulation of this experiment with the same dephasing values as in (b) is shown in (d).

We model the system in Fig. 1a with 4 quantum states: $|1\rangle$, $|2_X\rangle$, $|2_Y\rangle$, and $|3\rangle$. Two ac laser fields with $\pi_y$ polarization couple states $|1\rangle$ to $|2_Y\rangle$ and $|2_Y\rangle$ to $|3\rangle$ at angular frequencies $\omega_1$ and $\omega_2$, respectively. A master equation for the density matrix includes four decay terms which account for spontaneous emission: $\hbar\gamma_{32x} = \hbar\gamma_{32y} = 0.74$ µeV and $\hbar\gamma_{21x} = \hbar\gamma_{21y} = 1.13$ µeV. We note that coupling from $|3\rangle$ to $|1\rangle$ is dipole forbidden ($\hbar\gamma_{31} = 0$). This is crucial for observing quantum interference effects in a ladder system. We take the steady-state limit to describe the experiment as the integration time (time constant ≥ 1 s) is longer than the relevant QD dynamics. The experimental observables are the transmission and reflection signals, which are proportional to the susceptibility, equivalently an off-diagonal component of the density matrix [18]. The computed differential transmission or reflection signal is also dependent on a prefactor $\alpha_0$, which accounts for the oscillator strength, the laser spot size, wavelength, and refractive index [21]. Furthermore, $\alpha_0$ is influenced by the experimental geometry and spectral fluctuations. Figures 3b and 3d show simulations for the probe field reflection signal as a function of $\hbar\Omega_C$ and detuning $\delta_C$. To account for spectral fluctuations, we convolute the calculated spectrum with a Lorentzian function corresponding to the experimentally measured linewidth (FWHM). The prefactor $\alpha_0 = 0.03$ is determined from the probe differential reflection signal when $\hbar\Omega_C = 0$ and $\hbar\Omega_P = 0.4$ µeV using a 3 µeV Lorentzian convolution. Using these parameters, the model reproduces the experimental signal amplitude and energy splittings of Fig. 3a and c.

Figure 4 shows that the peak to peak splitting increases linearly with the strength of the coupling field. By swapping the coupling and probe fields, we have also observed the dressed states of the strongly driven $|1\rangle$ - $|2_Y\rangle$ transition. Notably, the ratio of peak splitting for the two pump / probe geometries is consistent with that expected from the direct lifetime measurements. These results demonstrate an elegant method to manipulate the transition energies of our solid-state nanostructure optically. This is increasingly important for applications. For example a strong coupling field far from resonance (ac Stark effect) can be used to tune transitions in QD molecules independently [22, 23], eliminate the fine-structure splitting of the single exciton for entangled photon generation [24], and to fine-tune a transition resonance relative to a cavity-mode for cavity QED [25].

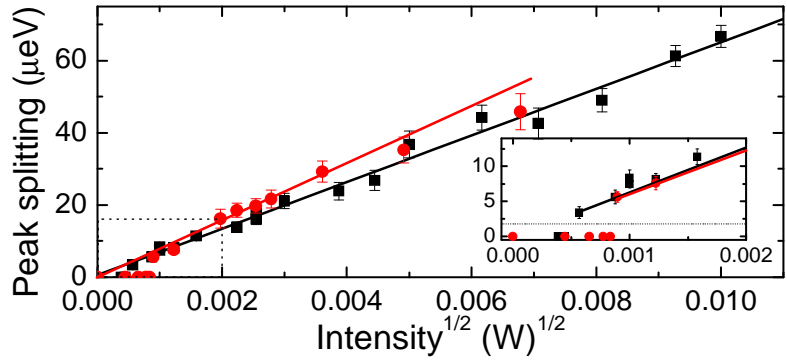

Figure 4: The peak to peak splitting, from ~ 3 to 66 µeV, varies linearly with the coupling field amplitude. The black squares (red circles) represent the peak splitting observed when the dressed states of the $|2_Y\rangle$ -$|3\rangle$ ($|1\rangle$ -$|2_Y\rangle$) are probed. The straight lines are fits to the data. For the fit of the red circles, the highest two intensity points are not taken into account as they showed anomalous features in the spectra. The inset highlights the data in the low saturation regime. The dashed line in the inset corresponds to minimum linewidth observed when $\hbar\Omega_C = 0$.

While the linear dependence of the Autler-Townes splitting persists to very large coupling field amplitudes ($\hbar\Omega_C \gg \hbar\gamma_{ij}$), in the weak field regime ($\hbar\Omega_C \approx \hbar\gamma_{ij}$) the peak splitting becomes obscured by the combined homogeneous and inhomogeneous contributions to the linewidth. The inset of Fig. 4 highlights the data in this regime. At the smallest intensities no splitting can be observed. However, the data show that the pump-probe geometry is crucial: a minimum splitting of 3.6 (5.6) µeV is distinguishable when the coupling field is resonant with the upper (lower) transition. This difference in the two pump-probe geometries is obvious in the numerical simulations shown in Fig. 5a and f. The parameters for $\gamma_{21}$, $\gamma_{32}$, $\hbar\Omega_P$, and Lorentzian broadening are the same as those defined for Fig. 3. In the case where $\Omega_C$ is applied to the upper states and the coherence of the lower states is probed (Fig. 5a), two peaks are distinguishable even when 0.71 $\hbar\Omega_C$ is smaller than the inhomogenously broadened linewidth (3 µeV), a strong indication of *destructive* quantum interference. In the simulation of the opposite pump / probe geometry (Fig. 5f), there is zero probe absorption signal when $\hbar\Omega_C = 0$ as the population resides in the ground state, $|1\rangle$. The signal then increases as $\hbar\Omega_C$ is increased until a maximum, ~ 10% of the maximum signal strength in Fig. 5a, is reached before the line begins to split into two peaks. In this simulation, two distinct peaks do not appear in the spectra until 0.71 $\hbar\Omega_C$ ~ 5 µeV, a strong indication of *constructive* quantum interference.

The remaining panels in Fig. 5 show the experimental (data points) and simulated (solid curves) evolution from a single, flat-topped peak into two distinct peaks as $\hbar\Omega_C$ is increased for both pump / probe geometries. The experimental spectra show quantitative agreement with the simulated spectra both in peak splitting and overall amplitude. A direct experimental comparison of the two pump / probe geometries can be made for the same coupling energies, $\hbar\Omega_C = 4.8$ µeV, in Figs. 5d and 5i. For this coupling energy, two distinct peaks are observed when the upper transition is strongly pumped and the lower transition probed. Conversely, only one flat-topped peak is visible when the pump and probe lasers are swapped. In this case, when the coupling energy is increased to $\hbar\Omega_C$ = 7.8 µeV the peak splitting can be resolved (Fig. 5h).

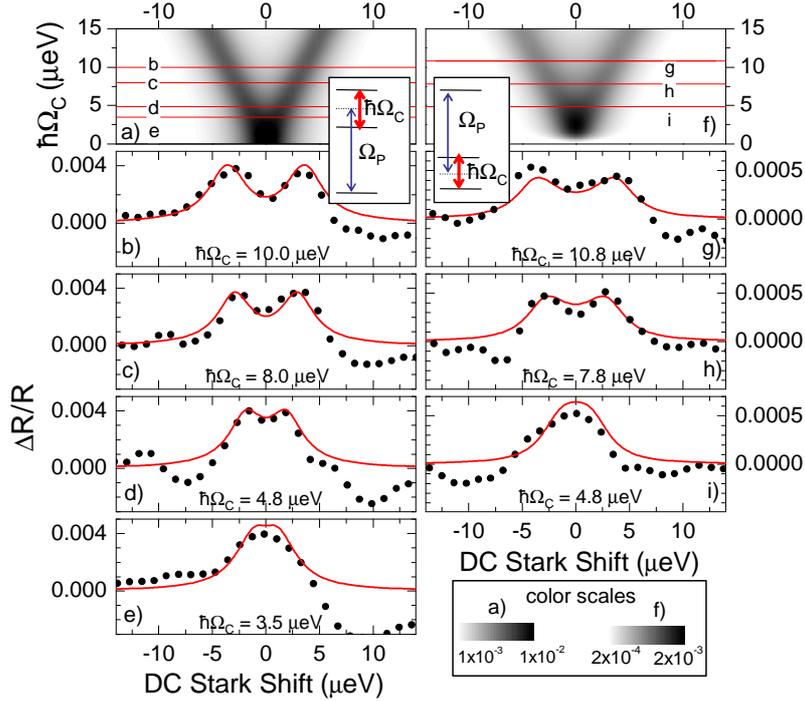

Figure 5: A comparison of the peak splitting and signal contrast in the low saturation regime for the two experimental geometries presented in Figs. 1b and 1c. The left hand column (panels a – e) corresponds to strongly pumping the upper transition and probing the lower transition; in the right hand panels (f – i) the coupling and probe lasers are swapped. The simulations shown in (a) and (f) highlight the different behavior: in (a) the peak splitting is distinguishable for smaller values of $\hbar\Omega_C$ than in (f). The grey scales have the dimensionless units $\alpha_0 \Delta R/R$. The experimental spectra in the low coupling power regime match that predicted by the numerical simulation. The striking difference of the two experimental pump / probe geometries can be made by comparing the data for $\hbar\Omega_C$ = 4.8 µeV and $\hbar\Omega_P$ = 0.4 µeV in (d) and (i) and the predicted spectra (red curves). The peak splitting is distinguishable in (d) whereas a flat-top, non-Lorentzian lineshape is measured in (i). The model quantitatively predicts both the lineshapes and signal amplitudes for each spectrum. The undershoot in the spectrum at ~10 µeV of (c) to (e) is due to a wavelength dependent interference effect in the reflectivity experiment.

We propose that the origin of the different behaviour at low pump power is a manifestation of quantum interference [1, 26]. In the first case, pumping the $|1\rangle$ - $|2_Y\rangle$ transition, there is an incomplete constructive interference; in the other case, pumping the $|2_Y\rangle$ - $|3\rangle$ transition, there is an incomplete destructive interference. Such effects in a ladder system are considered by Agarwal [26]. The probe field experiences an absorptive and a dispersive resonance at each dressed state and the net absorption spectrum can be constructed by summing the two absorptive and two dispersive contributions [26]. Significantly, the prefactor of the two absorptive contributions are always positive whereas the prefactor of the two dispersive components can be positive or negative depending on the pump / probe geometry and dephasing rates. Quantum interference takes place between the two absorption channels: in this picture a negative (positive) dispersive component for zero probe detuning results in destructive (constructive) interference. For further insight, Agarwal analytically solves for the absorption at the bare transition energy in the limit that $\hbar\Omega_C \gg \hbar\gamma_{ij}$ and $\hbar\Omega_P \ll \hbar\gamma_{ij}$. In this regime, the quantum interference can be characterized by the parameter β [26]. For the ladder system, in the limit where the non-radiative dephasing rates of levels $|2_Y\rangle$ and $|3\rangle$ are zero, $\beta = \gamma_{21} - \gamma_{32}$ for strongly pumping the upper transition and probing the lower transition and $\beta = -\gamma_{21}$ for strongly pumping the lower transition and probing the upper tranistion.

When the pump is resonant with the upper transition and $\gamma_{32} < \gamma_{21}$, β is positive. This is the situation for the QD studied here. A positive β signifies destructive quantum interference and the dispersive components are negative at the bare probe resonance. This situation is analogous to the prototypical "lambda" system which is commonly used for EIT [1]. In an idealized limit where state $|3\rangle$ is metastable (i.e. $\gamma_{32} \to 0$), the dispersive contributions exactly cancel the absorptive components and the probe absorption is *completely* cancelled. As the coherence of $|3\rangle$ is hypothetically shortened (i.e. $\gamma_{32}$ approaches the value $\gamma_{21}$), β approaches zero denoting that the interference effect is lessened and the probe absorption reappears. Conversely, for $\gamma_{32} > \gamma_{21}$, β is negative which signifies constructive quantum interference. In this scenario the dispersive components add to the absorptive con-tributions and the probe absorption is enhanced for zero probe detuning. While the analytical solution is valid within certain limitations, numerical simulations can include the exact experimental and QD parameters. Fig. 6 shows the result of numerical simulations for hypothetically varying β. In Fig. 6a, probe absorption spectra are displayed for three values of $\hbar\gamma_{32y}$ (0.06, 0.74, and 2.20 μeV) using the same QD and experimental parameters as Fig. 5a, confirming the interpretation of ref. 26 for this experiment. Conversely, when the pump and probe fields are swapped, β is always negative and the dispersive components are always positive at the bare state resonance. This leads to constructive interference and is analogous to the "V" system. Hence, rather than observing a dip in the probe absorption spectrum, only one flat-top peak is expected, even if state |2> is very coherent. This effect is simulated in Fig. 6b for $\hbar\gamma_{21}$ = 1.12, 0.13, and 6.58 μeV.

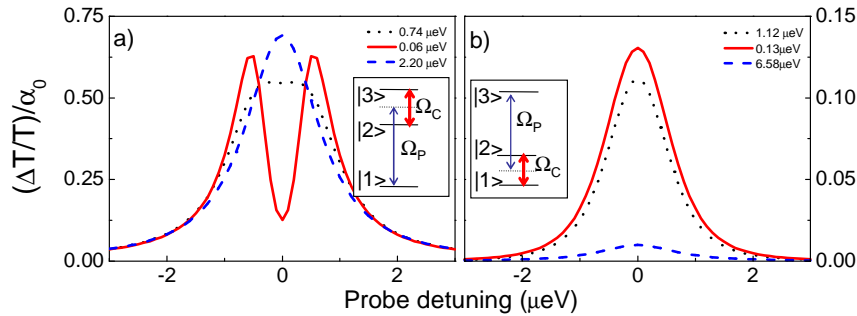

Figure 6: (a) The calculated effect of varying $\hbar\gamma_{32y}$ on the probe absorption spectrum when the upper ladder transition is strongly pumped. The following parameters are used: $\hbar\gamma_{21}$ = 1.12 μeV, $\hbar\Omega_P$ = 0.4 μeV, and $\hbar\Omega_C$ = 1.0 μeV. $\hbar\gamma_{32y}$ is listed in the legend. As $\hbar\gamma_{32y}$ increases the quantum interference changes from destructive to constructive and the dip at zero probe detuning disappears. (b) The probe absorption spectrum when the lower ladder transition is pumped. The following parameters are used: $\hbar\gamma_{32y}$ = 0.74 μeV, $\hbar\Omega_P$ = 0.4 μeV, $\hbar\Omega_C$ = 1.0 μeV, and $\hbar\gamma_{21}$ is listed in the legend. The dotted black lines show the conditions for the QD parameters in our sample.

The spontaneous emission rates in the QD are determined by the transition matrix element and the photon density of states. Our QD sample is in free space, hence there is a continuum of available photon modes. However, by incorporating QDs into micro-cavities the photon modes become discrete and modification of the spontaneous emission rate for different states in a QD becomes feasible [27, 28]. This technology offers a direct route to control β and thus modify both the visibility and nature (i.e. constructive or destructive) of quantum interference effects for the ladder system in a QD. In the current conditions (dotted black lines in Fig. 6), weak destructive (constructive) interference effects are observed when strongly pumping the upper (lower) transition and probing the lower (upper) transition. Notably, a 10-fold decrease in $\gamma_{ij}$ is possible with current technology [27, 28]; this would allow for much stronger interference effects to be manifest in a QD ladder system (solid red curves in Fig. 6).

In summary, we have observed the Autler-Townes splitting using both possible pump / probe geometries in a QD ladder system. Furthermore, our results confirm that modest quantum interference effects are present in this system. In higher dimensional structures such as quantum wells, coherence and quantum interference effects in three level ladder systems have also been observed [29, 30]. In these systems, the dephasing rates are ~ ps$^{-1}$ [29, 30], compared ~ ns$^{-1}$ dephasing rates in QDs. For the QD three level ladder system, quantum interference between two absorption channels is clearly observed but the effect has modest consequences owing to the slightly smaller dephasing from state |3> compared to |2> due to spontaneous emission. This suggests that striking quantum interference phenomena are achievable in a QD which is embedded in a micro-cavity. In this case both the strength and nature of the quantum interference become tunable.

Acknowledgements: We would like to thank F. Zimmer and P. Öhberg for fruitful discussions. This work was funded by EPRSC, Nanosystems Initiative Munich, and SANDiE. B.D.G. thanks the Royal Society of Edinburgh for financial support.